\documentclass[slac_one]{revtex4}
\usepackage{graphicx}
\usepackage{fancyhdr}
\begin{document}
\title{Search for long-lived particles in ATLAS and CMS}

\author{S.Giagu \\
{\it on behalf of the ATLAS and CMS collaborations}}
\affiliation{Sapienza Universit\`a di Roma and INFN Roma, 00185-Roma, IT}

\begin{abstract}
The ATLAS and CMS detectors can be used to search for heavy long-lived particles which might signal physics beyond the 
Standard Model. Such new states can be distinguished from Standard Model particles by exploiting their unique signatures, ranging from
multi-leptons and/or jets production anywhere within the detector volume, to minimum ionizing particles with low velocity and high momentum. 
Here are reviewed the strategies proposed by ATLAS and CMS to search for these signals, with particular emphasis on 
possible challenges to the trigger and detector operations. 
\end{abstract}

\maketitle

\thispagestyle{fancy}

\section{CHARGED META-STABLE PARTICLES}
Heavy stable charged particles are predicted by many models of physics beyond the Standard Model~\cite{ExpLim}, in which one or more new states carry a new conserved, or almost conserved, global quantum number, like for example R-parity or KK-parity.  
Stable heavy charged sleptons appear for example in Gauge Mediated Supersymmetry Breaking (GMSB) models,  when the lightest supersymmetric particle (NLSP), typically the stau slepton, couples weakly with the gravitino LSP.
Production of sleptons at the LHC proceeds mainly via decay chains of heavier supersymmetric particles, with typical cross-section at LHC ranging from 100 fb to 1 pb depending on the mass of the slepton particle.
Stable leptons are also predicted by the Universal Extra Dimensions model (UED), where for each Standard Model particle exists a corresponding so-called Kaluza-Klein (KK) state in extra dimensions, with the same quantum numbers and spin as the Standard Model partner.
KK leptons are directly produced in pairs in p-p collisions, with cross-section at LHC of the order of 20-30 fb.
In addition to lepton-like particles, also hadron-like stable particles (R-hadrons) can be produced at LHC. R-hadrons are predicted in Split-SUSY models, or in the framework of the gravitino LSP scenario of SUGRA models, and can be copiously produced at the LHC, with cross-sections up to few nb.
A common and largely model independent signature for all these processes is the existence of one or more stable massive charged particles, with low velocity and high transverse momentum, that behave similarly to "massive" muons in the LHC detectors. 

\subsection{TRIGGER}
At trigger level a lepton-like heavy stable particle has a high probability of being reconstructed as a muon. However particles with velocity significantly smaller than the speed of light may reach the muon system out of time with respect to the typical relativistic muons and therefore would either be reconstructed in the wrong bunch crossing or fail to be reconstructed at all because of quality cuts imposed by the Level-1 or High Level Trigger algorithms. This effect is more important for the ATLAS detector due to the larger dimensions compared to CMS.  Typically in GMSB models the two sleptons are produced with different velocities ($\beta$), and with the $\beta$ distribution peaking at high values, so that at least one of the two produced sleptons has $\beta > 0.7$ in 99\% of the events.  R-hadrons instead are much more problematic due to a much lower average $\beta$,  and the possibility of experiencing multiple nuclear interactions with the detector materials, with consequent charge flipping, and the chance to fail the quality requirements usually applied at high level of trigger.

\begin{figure*}[h]
\centering
\includegraphics[width=9.6cm]{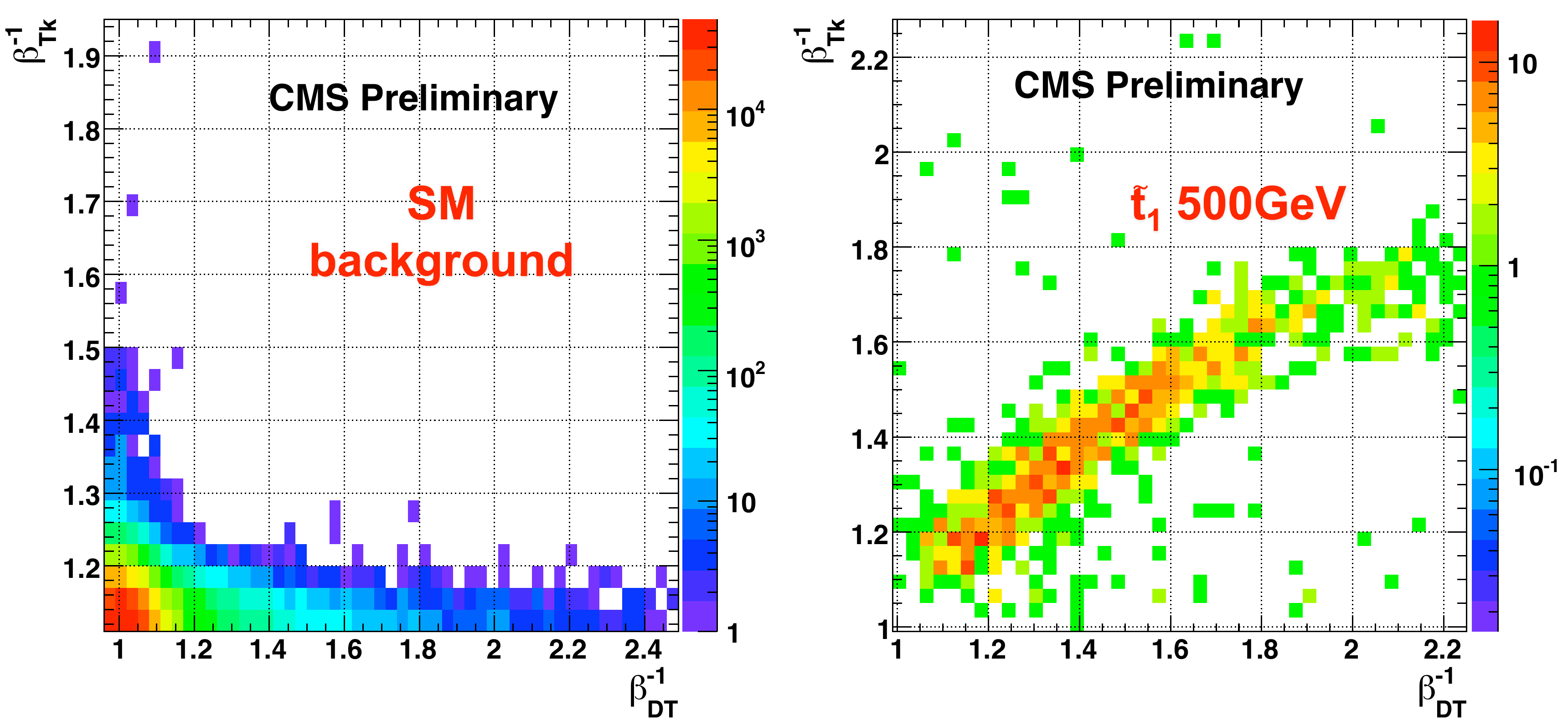}
\caption{CMS: distribution of  $\beta^{-1}$ as measured from $dE/dx$ (y axis) and from time-of-flight (x axis), for background (left) and for the signal sample $\tilde{t}$ (right).}
 \label{fig:rh3}
\end{figure*}

\subsection{VELOCITY MEASUREMENT}
The identification of a heavy stable charged particle relies on the precise determination of its mass through the measurement of both the velocity 
$\beta$ and the particle momentum. ATLAS and CMS are able to measure with precision $\beta$ using time-of-flight techniques, exploiting the excellent time resolution of the muon systems.
This can be done offline and also at trigger level, exploiting the fast time response of the Resistive Plate Chamber systems, available in both ATLAS and CMS experiment.
In addition to the time-of-flight determination, CMS performs also an independent $\beta$ measurement exploiting the specific ionization ($dE/dx$) measured in the central tracking system. The usage of $dE/dx$ allows also to reduce to a negligible level the background due to cosmic muons or to muons from different bunch crossings.  Figure~\ref{fig:rh3} shows the distribution of $\beta^{-1}$ as measured from  $dE/dx$ versus the determination from time-of-flight, for muon background events and for a possible R-hadron signal (stop squark of 500 GeV mass). A clear correlation between the two measurement is visible in the case of the signal.

\subsection{DISCOVERY POTENTIAL}
\begin{figure*}[h]
\centering
\includegraphics[width=5cm]{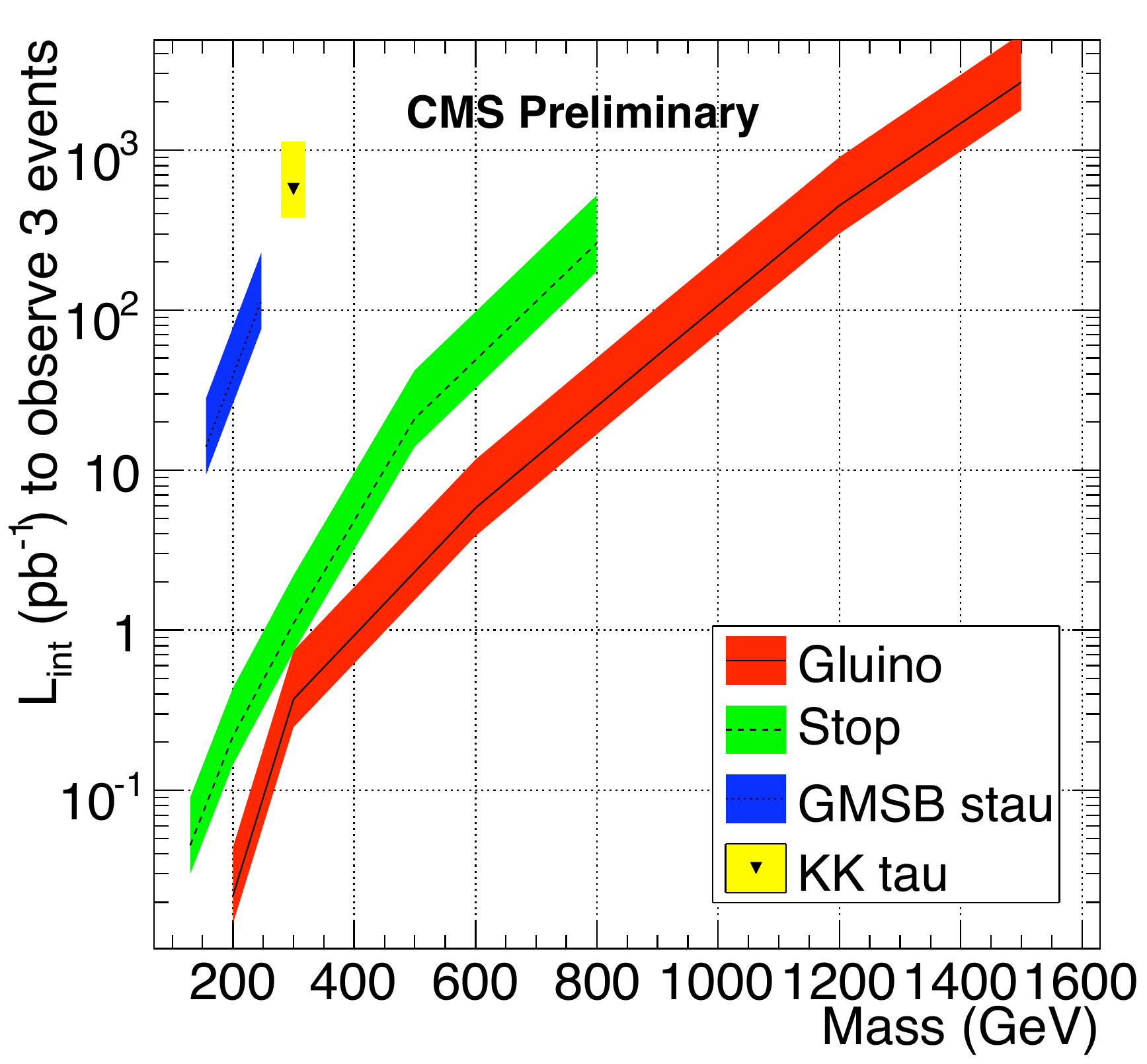}
\caption{CMS: integrated luminosity (pb$^{-1}$) needed to observe 3 signal events, for the four signal models (gluino, stop, KK $\tau$ and stau) as a function of the 
particle mass. The error bands correspond to a systematic uncertainty of 50\% on the estimated trigger efficiency.} 
 \label{fig:rh4}
\end{figure*}

In order to maximize the background rejection, CMS uses a combined selection based on the two measurements of $\beta$, reducing the background to a negligible level without significant loss of signal.  Using this technique CMS is able to keep the expected background below one event for an integrated luminosity of 1 fb$^{-1}$, maintaining reasonable signal efficiencies.   
The CMS integrated luminosity needed to observe 3 events in the signal region, for different models is shown in Figure~\ref{fig:rh4}.  Heavy Stable Charged Particles can be discovered with early data for different models and in different mass regions. The stable gluino search with 1 fb$^{-1}$ is sensitive to gluino masses above 1 TeV and the GMSB scenarios with stable stau can be discovered with a few 100 pb$^{-1}$~\cite{cms_rh}. 
Similar sensitivities are obtained also by the ATLAS collaboration~\cite{atlas_rh}. 
\section{NEUTRAL LONG-LIVED PARTICLES} 
A number of extensions of the Standard Model result in particles that are neutral, weakly-coupled, and have macroscopic decay lengths that can be comparable with LHC detector dimensions. A large class of such models are represented by the Hidden Valley models~\cite{hv1}, where the SM is extended  by a hidden sector, the Òv-sectorÓ for short, and a communicator (or communicators) which interacts with both sectors. A barrier (perhaps the communicatorÕs high mass, weak couplings, or small mixing angles) weakens the interactions between the two sectors, making production even of light v-sector particles (Òv-particlesÓ) rare at low energy. At the LHC, by contrast, production of v-particles, through various possible channels, may be observable. The communicator can be any neutral particle or combination of particles, including the Higgs boson, the Z' bosons, or the SUSY LSP for example.
Here we present the results of a first study of the ATLAS detector performance for the Higgs decay 
$h^0\to \pi_v\pi_v$ , where $\pi_v$ is a new massive neutral particle with long lifetime (1.5 m in this study), that decay mainly in a pair of bottom quarks. 

\subsection{DETECTOR SIGNATURES AND TRIGGER SELECTION}
\begin{figure*}[h]
\centering
\includegraphics[width=5.5cm]{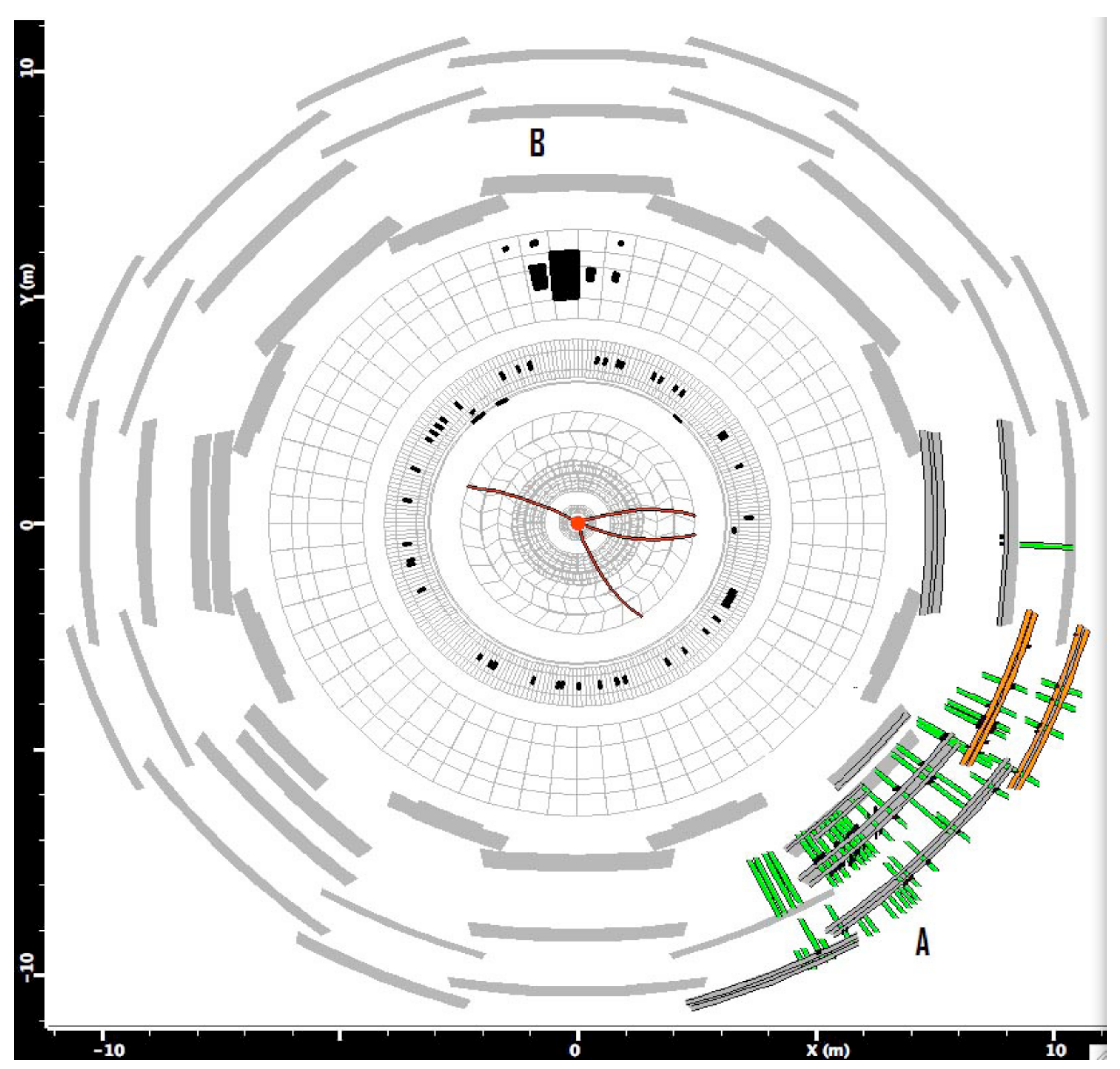}
\caption{ATLAS event display for the process $h\to\pi_v\pi_v$. One of the $\pi_v$ decays in the muon spectrometer, while the other decays in the hadron calorimeter.}
 \label{fig:hv1}
\end{figure*}
\begin{figure*}[h]
\centering
\includegraphics[width=10.4cm]{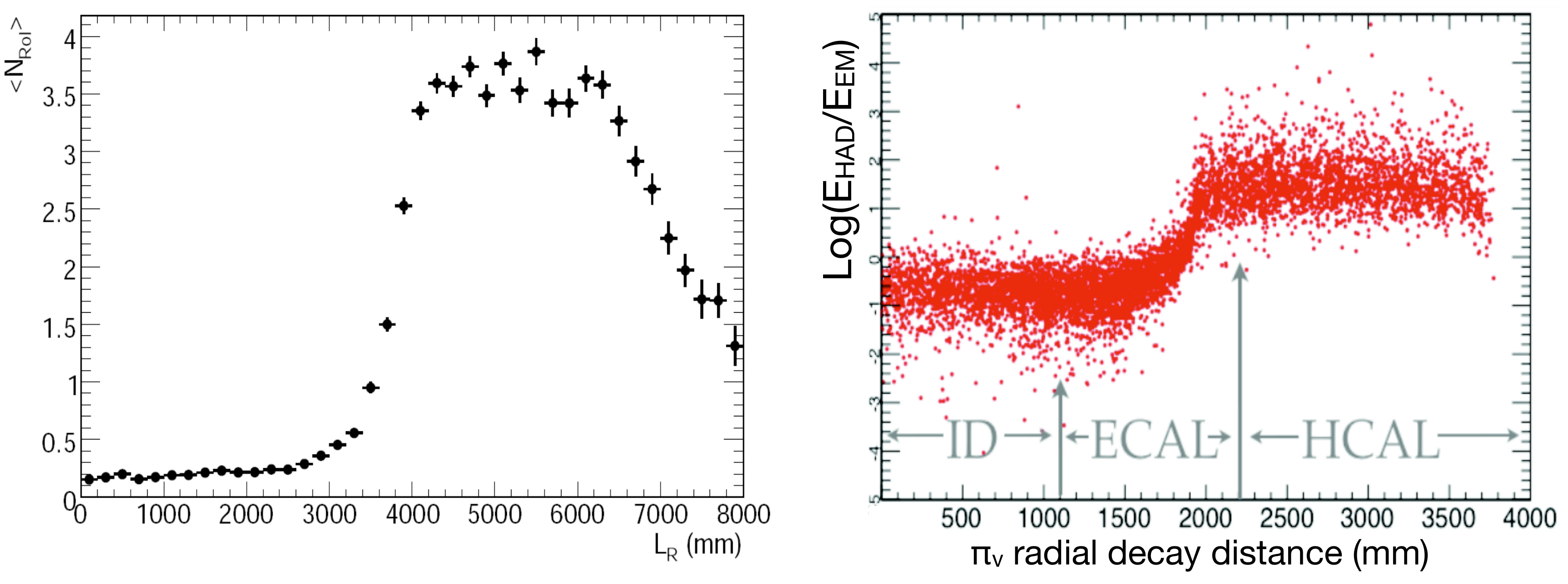}
\caption{ATLAS: Left:  average number of Level-1 muon candidates contained in a cone of $\Delta R=0.4$  around the $\pi_v$ line of flight vs $\pi_v$ radial decay distance. Right:  $\log_{10} (E_{HAD} /E_{EM} )$ vs $\pi_v$ decay distance.}
 \label{fig:hv2}
\end{figure*}

A simulation of a typical $h^0\to\pi_v\pi_v\to b\bar{b}b\bar{b}$ event in the ATLAS detector is shown in Figure~\ref{fig:hv1}. Due to the displaced vertices with tracks non pointing to the interaction region, and to the low Higgs mass (140 GeV in this simulation), the standard ATLAS triggers~\cite{atlas_trig} are able to select only a very small fraction of these events (typical  Level-1 trigger efficiencies are smaller than 5\%).
A signature driven trigger strategy is therefore required. We focused on two detector regions to illustrate the trigger signatures of Hidden Valley particles: decays in the muon spectrometer, and decays in the calorimeters. These are both the most challenging from the trigger point of view and
the ones giving the most striking experimental signatures for a possible early discovery.
Decays occurring near the end of the hadron calorimeter and before the first muon trigger plane result in a large number of hadrons traversing a narrow ($\eta,\phi$) region of the muon spectrometer. The Level-1 muon trigger will return several clustered muon candidates, as shown in Figure~\ref{fig:hv2}, where the Level-1 muon candidates contained in a cone of radius $\Delta R = 0.4$ around the $\pi_v$ line of flight, as a function of the $\pi_v$ radial decay distance, are plotted. As the $\pi_v$ decay vertex approaches the end of the hadron calorimeter (4500 mm), the average number of muon candidates contained in the cone plateaus at $\sim$3.5 until the $\pi_v$ decays close to the first trigger plane (7000 mm), at which point the charged hadrons are not spatially separated enough to give multiple unique muon candidates. 
This signature can be used as a stand-alone Level-2 trigger object to select with good efficiency these late decays. 
Events with decays near the outer edge of the electromagnetic calorimeter and in the hadron calorimeter, are characterized by jets with few or no tracks and unconventional energy distributions (jets with more energy deposited in the hadron calorimeter than in the electromagnetic one). The logarithm of the hadronic to electromagnetic energy ratio for jets from $\pi_v$ decays as a function of 
the $\pi_v$ decay distance can be seen in Figure~\ref{fig:hv2}.  As the $\pi_v$ decays closer to the end of the electromagnetic calorimeter  (2200 mm), the ratio changes from a  characteristic negative to a positive value. 
Displaced decays in the outer part of the inner detector or inside the electromagnetic calorimeter result in low tracking efficiency, because tracking requires seed hits in the pixel and silicon strip layers. This suggests that a jet with no tracks reconstructed in the inner detector may be used to select 
$\pi_v$ decays in the electromagnetic calorimeter. In this case, to reduce QCD background, a Level-1 muon candidate contained in a cone of radius $\Delta R=0.4$ around the jet axis is required, which selects a semileptonic decay of one of the two b daughter of the $\pi_v$ particle. 
\subsection{TRIGGER PERFORMANCE}
In Table~\ref{tab:hv1} the ATLAS trigger acceptances for the new signature based triggers described above are shown. With these new 
triggers ATLAS will be able to select  about 20\% of events with displaced decays from $h^0\to\pi_v\pi_v$.
Standard Model QCD processes are a potential source of significant background at the trigger level. The same trigger objects have been applied to fully simulated di-jet samples, resulting in a negligible (6 nb) cross-section acceptance at Level-2.  In conclusion long lived neutral particles predicted by Hidden Valley models, can be successfully collected by implementing dedicated signature based triggers, that allow to increase the selection efficiency with a negligible background rate from Standard Model processes. 
\begin{table}[h]
\begin{center}
\caption{Hidden Valley specific triggers efficiency, normalized to the whole sample.}
\begin{tabular}{|c|c|c|c|c|}
$\log_{10} (E_{HAD} /E_{EM})$ & Trackless Jet & Muon Cluster & Total HV Triggers & Total All Triggers \\\hline
5\% & 3.8\% & 9\% & 15.7\% & 18.5 \% \\\hline
\end{tabular}
\label{tab:hv1}
\end{center}
\end{table}
\vspace{-0.3cm}
\begin{acknowledgments}
The author wish to thank Matt Strassler (Rutgers University) for invaluable contributions in many stages of the work related to the Hidden Valley scenario.
\end{acknowledgments}

\end{document}